\documentclass[aps,twocolumn,amssymb,superscriptaddress]{revtex4}

\usepackage{amsmath}    \usepackage{latexsym}    \usepackage{graphicx}
\usepackage{times}

\begin{document}

\title{FRAGMENTATION OF GENERAL RELATIVISTIC QUASI-TOROIDAL POLYTROPES}

\author{Burkhard~Zink} 
\affiliation{Center for Computation and
Technology, Louisiana State University, Baton Rouge, LA 70803, USA}
\affiliation{Horace Hearne Jr. Institute for Theoretical Physics,
Louisiana State University, Baton Rouge, LA 70803, USA}

\author{Nikolaos~Stergioulas}   \affiliation{Department of Physics,
Aristotle University of Thessaloniki, Thessaloniki 54124, Greece}

\author{Ian~Hawke}  \affiliation{School of Mathematics,  University of
Southampton, Southampton SO17 1BJ, UK}

\author{Christian~D.~Ott} \affiliation{Department of Astronomy and Steward Observatory,
The University of Arizona, Tucson, AZ, USA}

\author{Erik~Schnetter}   \affiliation{Center for Computation and
Technology, Louisiana State University, Baton Rouge, LA 70803, USA}
\affiliation{Max-Planck-Institut f\"ur
Gravitationsphysik, Albert-Einstein-Institut, 14476 Golm, Germany}

\author{Ewald~M\"uller}     
\affiliation{Max-Planck-Institut f\"ur Astrophysik,
Karl-Schwarzschild-Str.\ 1, 85741 Garching bei M\"unchen, Germany}

\maketitle

How do black holes form from relativistic stars? This question is
of great fundamental and practical importance in gravitational physics
and general relativistic astrophysics. On the fundamental level, black
holes are genuinely relativistic objects, and thus the study of their
production involves questions about horizon dynamics, global structure
of spacetimes, and the nature of the singularities predicted as a 
consequence of the occurrence of trapped surfaces.  On the level of
astrophysical applications, systems involving black holes are possible
engines for highly energetic phenomena like AGNs or gamma-ray bursts,
and also likely a comparatively strong source of gravitational radiation.

The most simple model of black hole formation from, say, cold neutron stars,
is a fluid in spherically symmetric polytropic equilibrium moving on a 
sequence of increasing mass due to accretion \cite{Shapiro83}. This assumes that 
(i) the stellar structure and dynamics are represented reasonably by the
ideal fluid equation of state and the polytropic stratification, (ii) accretion
processes are slow compared to the dynamical timescales of the star, and 
(iii) rotation is negligible. Our focus has been to study the effects of 
relaxing the third assumption.

In spherical symmetry, the sequence of equilibrium polytropes has a 
maximum in the mass function $M(\rho_c)$, where $\rho_c$ denotes the central
rest-mass density of the polytrope. This maximum is connected to a change
in the stability of the fundamental mode of oscillation \cite{Shapiro83}, and thus
collapse sets in via a dynamical instability to radial deformations. During
the subsequent evolution, a trapped tube forms at the center which traverses the stellar
material entirely \cite{Shapiro80}.

How much of this behaviour is preserved when rotation is taken into account?
Rotation is known to change the equilibrium structure of the star, and, in
consequence, its modes of oscillation and set of unstable perturbations. The collapse
might also lead to the formation of a massive disk around the new-born black hole, and
finally only systems without spherical symmetry can be a source of gravitational
radiation. 

Numerical simulations have been used to study the collapse and black hole formation of 
general relativistic rotating polytropic stars \cite{Baiotti04}. For the uniformly and
moderately differentially rotating models investigated in those studies, the dynamical 
process is described by the instability of a quasi-radial mode and subsequent collapse
of the star up to the formation of an accreting Kerr black hole at the star's center. 

Will strong differential rotation modify this picture? Even before our study, there was
evidence that this should be the case. (i) Strong differential rotation can deform the
high-density regions of a star into a \emph{toroidal} shape, thus changing the equilibrium
structure considerably. (ii) It admits stars of high normalized rotational energy $T/|W|$ \cite{Shapiro83}
which are stable to axisymmetric perturbations. (iii) It admits non-axisymmetric instabilities, for
example by the occurrence of corotation points\cite{Watts:2003nn}, at low values of 
$T/|W|$\cite{Centrella:2001xp}. (iv) A bar-mode instability of the type found in Maclaurin
spheroids\cite{Chandrasekhar69c} would likely express itself by the formation of two orbiting 
fragments if the initial high-density region has toroidal shape.

This last property has motivated us to ask this question: \emph{Can a bar deformation transform
a strongly differentially rotating star into a binary black hole merger with a massive accretion
disk?} If so, this process might occur in supermassive stars if the timescales associated with 
angular momentum transport are too large to enforce uniform rotation.

\begin{figure}
\begin{center}
\includegraphics[width=0.6\columnwidth]{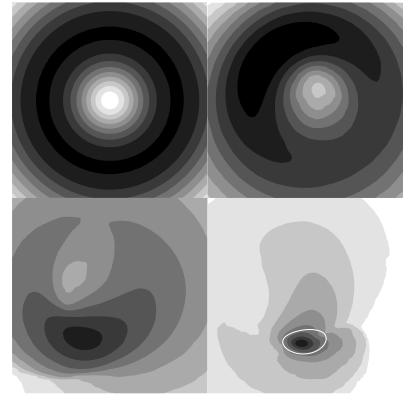}
\end{center}
\caption{Development of the fragmentation instability in a model of a strongly differentially 
rotating supermassive star. The darker shades of grey indicate higher density. The closed
white line in the last plot is a trapped surface.}
\end{figure}

We have investigated black hole formation in strongly differentially rotating, quasi-toroidal
models of supermassive stars \cite{Zink2005a, Zink2006a}, and found that a non-axisymmetric 
instability can lead to the off-center formation of a trapped surface (see figure). An extensive
parameter space study of this fragmentation instability \cite{Zink2006a} reveals that many 
quasi-toroidal stars of this kind are dynamically unstable in this manner, even for low values
of $T/|W|$, and we have found evidence that the corotation mechanism observed by 
Watts et al. \cite{Watts:2003nn} might be active in these models. Since, on a sequence of increasing
$T/|W|$, one of the low order $m=1$ modes becomes dynamically unstable before $m=2$ and higher order
modes, one would not expect a binary black hole system to form in many situations (although this may
depend on the rotation law and details of the pre-collapse evolution as well). Rather, the off-center
production of a single black hole with a massive accretion disk appears more likely.

Since the normalized angular momentum $J/M^2$ of the initial model is greater than unity, there is
another interesting consequence of this formation process: the resulting black
hole, unless it is ejected from its shell, may very well
be \emph{rapidly rotating}, spun up by accretion of the material remaining outside the initial location
of the trapped surface. Investigating the late time behaviour of this accretion process, estimating
possible kick velocities of the resulting black hole, and finding the mass of the final accretion disk
is, however, beyond our present-day capabilities and subject of future study.


\begin{thebibliography}{00}

\bibitem{Shapiro83} S. Shapiro and S. Teukolsky, {\it Black Holes, White Dwarfs
and Neutron Stars} (Wiley 1983).

\bibitem{Shapiro80} S. Shapiro and S. Teukolsky, {\it Astrophys. J.} {\bf 235}, 199
(1980).

\bibitem{Baiotti04} M. Shibata, T. Baumgarte and S. Shapiro, {\it Phys. Rev. D} {\bf 61},
044012 (2000). L. Baiotti, I. Hawke, P. Montero, F. L{\"o}ffler, L. Rezzolla,
N. Stergioulas, J. A. Font and E. Seidel, {\it Phys. Rev. D} {\bf 71}, 024035 (2005),
and references therein.

\bibitem{Watts:2003nn} A. Watts, N. Andersson and D. Jones, {\it Astrophys. J.} {\bf L37}
(2005).

\bibitem{Centrella:2001xp} J. Centrella, K. New, L. Lowe and J. Brown, {\it Astrophys. J.}
{\bf 550} (2001).

\bibitem{Chandrasekhar69c} S. Chandrasekhar, {\it Ellipsoidal Figures of Equilibrium}
(Yale UP 1969).

\bibitem{Zink2005a} B. Zink, N. Stergioulas, I. Hawke, C. D. Ott, E. Schnetter
and E. M\"uller, {\it Phys. Rev. Letters} {\bf 96}, 161101 (2006).

\bibitem{Zink2006a} B. Zink, N. Stergioulas, I. Hawke, C. D. Ott, E. Schnetter
  and E. M\"uller, astro-ph/0611601 (2006).

\end{thebibliography}
\end{document}